\documentstyle[11pt,newpasp,twoside]{article}
\def\edcomment#1{\iffalse\marginpar{\raggedright\sl#1\/}\else\relax\fi}
\reversemarginpar

\def\MSUN{\rm M_{\odot}}
 
\def\MSUNYR{\rm M_{\odot}\,yr^{-1}}

\newbox\grsign \setbox\grsign=\hbox{$>$} \newdimen\grdimen \grdimen=\ht\grsign
\newbox\simlessbox \newbox\simgreatbox
\setbox\simgreatbox=\hbox{\raise.5ex\hbox{$>$}\llap
     {\lower.5ex\hbox{$\sim$}}}\ht1=\grdimen\dp1=0pt
\setbox\simlessbox=\hbox{\raise.5ex\hbox{$<$}\llap
     {\lower.5ex\hbox{$\sim$}}}\ht2=\grdimen\dp2=0pt

\begin{document}
\title{Mass Outflows from Accretion Disks: Old and New Challenges}
\author{Daniel Proga}
\affil{JILA, University of  Colorado,  Boulder, CO 80309,  USA}

\begin{abstract}
We summarize recent developments in modeling of mass outflows from accretion
disks in AGN. We illustrate how the hydrodynamical and 
magnetohydrodynamical wind structures revealed in numerical simulations 
relate to observations. Important results from the numerical
simulations include:
no shielding from X-rays is needed to radiatively launch a wind from the inner
disk; shielding is needed to accelerate the wind to high velocities.
Disk winds launched magnetically seem to be too dense to
be accelerated by line driving. It appears that
we have an {\it all or nothing} situation for radiation-driving:
radiation driving has to be responsible for both launching and accelerating
the wind or a wind launched by some other mechanism must have 
a very finely tuned density in order  to be  accelerated by radiation driving
to velocities as high as measured in BAL QSOs. 

\end{abstract}

\section{Introduction}

Mass outflows in active galactic nuclei (AGN) are fairly common.
Broad absorption lines (BALs) in QSOs are perhaps the most compelling
evidence for such outflows. The BALs are always blueshifted relative to 
the emission-line rest frame,\footnote{A few SDSS BAL quasars have
a redshifted and a blueshifted component; see Hall et al.\ 2002.}
indicating the presence of outflows from 
the active nucleus, with velocities in the ultraviolet
as large as $66,000~\rm km\,s^{-1}$.
Other evidence for AGN outflows include narrow absorption lines (NALs).
NALs observed in ultraviolet (UV) observations of some QSOs
can be blueshifted by as much as $\sim~50,000~{\rm km~s^{-1}}$  
(Hamann et al.\ 1997). 
NALs observed in UV observations of many
Seyfert galaxies are blueshifted by several 
$100~{\rm km~s^{-1}}$ (Crenshaw et al.\ 1999).  
NALs can also be due to highly ionized species such as those observed in
a high-resolution X-ray observation of the Seyfert galaxy NGC~5548 
obtained by {\em Chandra} (Kaastra et al.\ 2000). We refer a reader to 
Arav, Shlosman, \& Weymann (1999); Crenshaw, 
Kraemer, \& George (2002); and 
these proceedings for more details  on observations 
of AGN outflows.

One of most plausible scenarios for AGN outflows is a wind from an
accretion disk around a black hole. However, even if we agree
that AGN outflows are indeed disk outflows, 
there are still many key questions  that
await answers. Here we mention  just five:
(i) From what radius on the disk do the outflows come?
(ii) Do the outflows need shielding from X-rays to avoid overionization?
(iii) What is the mechanism responsible for launching the outflows
from the disk? (iv) What is the mechanism responsible for accelerating
the outflows to their terminal velocities?
(v) All AGN are believed to have accretion disks, why then do not all
AGN exhibit outflows?

Our goal here is to review  various theoretical results
and try to identify a physical disk wind model which is most promising
in answering as many questions as possible concerning AGN flows
with as few assumptions as possible. 

Before we proceed, let us consider a specific question: 
is the terminal velocity 
of an outflow, $v_\infty$, a good indicator of the radius, $r_l$, 
from which the outflow was launched? One can argue that the answer to this
question is positive if the terminal velocity is correlated with the 
escape velocity, $v_{esc}$, from the location where the flow was launched. 
Such a correlation between $v_\infty$ and $v_{esc}$ is one
of the predictions of the radiation-driven wind theory 
(i.e., $v_\infty \approx~{\rm a~few}~v_{esc}$, see Proga 1999 and references therein).

\section{Hydrodynamical Models}

Radiation pressure on spectral lines (line force) driving
a wind from an accretion disk
is the most promising hydrodynamical (HD) scenario for AGN outflows.  
Within the framework
proposed by Murray et al.\ (1995) all five questions
mentioned above can be addressed  in one rather simple 
way: a wind is launched from the disk
by  the local disk radiation at radii where the
disk radiation is mostly in the UV. 
Such a wind  is continuous and requires shielding.
The UV radiation emitted inside the wind launching region
accelerates (radially) the wind.
Consequently, the wind is
equatorial and produces spectral features
only for certain (high) inclination angles.

In Proga, Stone, \& Kallman (2000, hereafter PSK), we attempted to
model line-driven winds in AGNs within this framework.
We applied line-driven stellar wind
models (Castor, Abbott, \& Klein 1975, hereafter CAK) to  winds driven 
from accretion disks. We also took into account
some of the effects of photoionization. In particular, we calculated the
gas temperature  assuming that the gas  is optically thin to its own
cooling radiation. 
In PSK, we took also into account 
the effects of photoionization on the line force by computing 
the parameters of the line force using a current value of the photoionization 
parameter, $\xi$, adopting results of Stevens \& Kallman (1990, hereafter SK). 
SK's results show that the line
force decreases sharply with increasing photoionization parameter.

In  PSK, we  found that a disk accreting onto a $10^8~\MSUN$ black hole 
at the rate of $1.8~\MSUNYR$ can launch a wind at $r_l\sim 10^{16}$~cm from 
the central engine. The X-rays from the central object are significantly 
attenuated by the disk atmosphere so they cannot prevent the local disk 
radiation from pushing matter away from the disk. 
However, in the supersonic portion of the flow high above the disk, 
the X-rays can overionize the gas and decrease the wind terminal velocity.
For a reasonable X-ray opacity, e.g., $\kappa_{\rm X}=40~\rm g^{-1}~cm^2$, 
the disk wind can be accelerated by the central UV radiation to velocities of 
up to 15,000~$\rm km~s^{-1}$ at a distance of $\sim 10^{17}$~cm 
from the central engine. We note that this velocity agrees very well
with the prediction of  the standard CAK model for the wind
terminal velocity (see e.g., Proga 1999).
The covering factor of the disk wind is $\sim 0.2$. The wind is unsteady
and  consists of an opaque, slow vertical flow near 
the disk that is bounded on the polar side by a high-velocity stream. 
A typical column density radially 
through the fast stream is a few $10^{23}~\rm cm^{-2}$ so the stream 
is optically thin to the UV radiation.  This low column density
is precisely why gas can be accelerated
to high velocities. The fast stream contributes nearly 100\% to 
the total wind mass loss rate of $0.5~~\MSUNYR$. 

In PSK, we took into account the attenuation of the X-ray radiation
by computing the X-ray optical depth in the radial direction and
assuming a simple formula for the $\kappa_{\rm X}$ dependence on the
photoionization parameter. 
It remains
to be seen whether fully self-consistent, multidimensional photoionization
and dynamical 
calculations will confirm that a wind launched from a disk by the local UV
disk radiation can shield itself from the X-rays 
in order to achieve velocities as high as those observed in AGN winds.
Such photoionization calculations in connection with two-dimensional, 
time-dependent HD calculations are just starting to become  feasible.

We emphasize that, within the framework of line driving,
X-ray shielding is necessary for an observer to see AGN outflows  
in UV absorption lines and for the outflows  
to be accelerated.
Assuming a steady state radial outflow and imposing total momentum
conservation,
one can show that unshielded gas outflowing with high velocities
has too high a photoionization parameter 
to produce UV lines and to be driven by the line force.
This result holds for an outflow 
consisting of a continuous wind as well as for 
an outflow consisting of  dense clouds. The latter  
needs shielding from X-rays if the filling factor decreases
with increasing radius --- a requirement we do not know how to satisfy
from first principles.
A need for some shielding of clouds 
has been already hinted at in the literature (e.g., dKB).

To avoid overionization without shielding, the wind which is launched  
by a force other than the line force must be
denser than the wind predicted by the line-driven wind models 
by several orders of magnitude. 
Thus if AGN outflows are unshielded 
and launched from an accretion disk at relatively large distances from 
the central object 
(where the local line force due to the disk is negligible) then we 
have to figure out not only what mechanism launches
this  dense outflow but also what mechanism accelerates
the outflow to velocities much  higher
than the escape velocity from the location of the launch.
Our conclusion that shielding of the wind from the X-rays
is required regardless  of the filling factor of the line-driven
outflow is consistent with 
the observed anti-correlation for QSOs between 
the relative strength of the soft X-ray flux and the CIV
absorption equivalent width (e.g., Brandt, Laor, \& Wills 2000).

The above conclusions are based on assumptions more simplified 
than those adopted
in our previous work (e.g., PSK, Proga \& Kallman 2002). Nevertheless they 
are consistent with numerous simulations.
Finally, we note that the wind terminal velocity
sets an upper limit for the radius from which the wind is launched,
provided the wind is line driven.
The line-driven wind is more likely to achieve
lower rather  than  much higher velocity compared to the escape velocity 
from the launching radius. The wind velocity can be lowered
if the wind is overionized downstream as in PSK's simulations [see
Proga (1999) for a discussion of when the wind terminal velocity 
can be lowered].
However, to accelerate radiatively a wind to velocities
higher than $v_{esc}$, the wind density would have
to be significantly lower than the density of the line-driven wind, but this
makes the wind very likely to be overionized and unable to
be accelerated at all.

One of most appealing aspects of line-driven disk wind
models is the fact that they predict from
first principles outflows that are consistent with observations.
Moreover, it is very encouraging
that line-driven models, when applied to disk winds from cataclysmic 
variables, are not only consistent with observations
but also are capable of reproducing them (e.g., Proga 2003b). 

\section{Magnetohydrodynamical Models}
Magnetically driven winds from disks are the favored explanation for 
the outflows in many astrophysical environments (e.g., 
Blandford \& Payne 1982; Uchida \& Shibata 1985).
Such winds do not require radiation pressure and thus can be important
in low luminosity  systems such as young stellar objects and
systems where gas can be overionized by very strong radiation
as in AGN. In the context of AGN outflows,
magnetically driven wind models usually rely on the effects of 
magnetic fields as well as on line driving.
For example, within the framework
proposed by de Kool \& Begelman (1995; see also Everett 2004, this 
volume) the five questions
mentioned in Section 1 can be addressed  in the following way: 
a wind is launched from an outer cold disk. 
Such a wind  is made of dense clouds  and does not require shielding.
The wind is accelerated  by the UV radiation emitted from the inner disk. 
The wind may be but does not have to be
equatorial and can produce spectral features
for all inclination angles provided there is at least one cloud
in the line of sight.

At the moment it is still unclear whether a purely line-driven
disk wind model or whether a hybrid model with magnetic and line-driving
better explains observations. One may prefer the former 
simply because it does not invoke processes that cannot be 
modeled from first principles. In the latter case, 
one has to postulate a mechanism responsible for launching
the clouds from the outer cold disks and a mechanism
to confine the clouds so they will not be destroyed or overionized.
However, this situation may change soon as new studies bring
new insights into
the physics of accretion disks and highly inhomogeneous
magnetized flows. 

For example, in Proga (2003a) we reported on numerical simulations of
the two-dimensional, time-dependent magnetohydrodynamics (MHD)
of  line-driven winds  from luminous accretion disks initially threaded
by a purely axial magnetic field. We focused on a generic
disk wind problem and did not include strong ionizing radiation.
Such studies can help to understand AGN outflows
because they address, for example, the problem of a wind
produced outside a shielding, inner outflow. 

In Proga (2003a) we used ideal MHD to compute 
the evolution of Keplerian disks, varying the magnetic field 
strength and the luminosity of the disk, the central accreting object, 
or both. 
We found  that the magnetic field very quickly starts deviating
from purely axial due to the magnetorotational 
instability. This leads to fast growth of the
toroidal magnetic field as field lines wind up due to the disk rotation.
As a result the toroidal field dominates over the poloidal field
above the disk and the gradient of the former drives a slow and dense disk 
outflow, which conserves the specific angular momentum of the fluid. 
Depending on the strength of the magnetic field relative to the 
system luminosity the disk wind can be radiation- or MHD-driven.
The pure line-driven wind consists of a dense, slow outflow that is 
bounded on the polar side by a high-velocity  stream. The mass-loss rate is 
mostly due to the fast stream. As the magnetic field strength increases first 
the slow part of the flow is affected, namely it becomes  denser and slightly 
faster and begins to dominate the mass-loss rate. In very strong magnetic 
field or pure MHD cases, the wind consists of only a dense, slow outflow 
without the presence of the distinctive fast stream so typical of pure 
line-driven winds. Our simulations indicate that winds launched by 
the magnetic fields are likely to remain dominated by the fields downstream 
because of their relatively high densities. Line driving 
may not be able  to change a dense MHD wind because the line force
strongly decreases with increasing density.

\section{Final Remarks}

Recent numerical simulations in the HD limit as well as in  the MHD limit
reveal many important features of disk outflows. For example,
no shielding is needed to radiatively launch a wind from the inner
disk. Shielding is needed to accelerate the wind to high velocities.
Disk winds launched magnetically seem to be too dense to
be accelerated by line driving. It appears then that
we have an {\it all or nothing} situation for line-driving:
line driving has to be responsible for both launching and accelerating
the wind; otherwise we require a rather fine tuning of the wind density 
so that some other mechanism produces an outflow with velocities
as measured in BALQSOs. All of this is because
the ionization balance is 
a key constraint on AGN outflows.

There are two groups modeling in great detail the ionization
structure of AGN outflows: (i) Everett (2004, this volume)
explores self-similar MHD disk wind models where an assumed steady
state enables a detailed treatment of the photoionization
and radiative transfer in one-dimensional approximation.
(ii) Chelouche (2001; 2003) considered purely radiation-driven  
one-dimensional
steady outflows. It is just a matter of time before multidimensional
time-dependent simulations with sophisticated photoionization calculations 
will be applied to AGN outflows
and we shall understand  AGN outflows as well as 
 winds in OB stars.

\acknowledgements
We acknowledge support from NASA under LTSA grants
NAG5-11736 and NAG5-12867.

\end{document}